\documentclass[prl,reprint]{revtex4-2}

\usepackage{appendix}
\usepackage{amsfonts,bm}
\usepackage{amsmath,amssymb,amsthm}

\usepackage{graphicx}
\usepackage{epstopdf}
\usepackage{dcolumn}
\usepackage{mathrsfs}
\usepackage{tikz}
\usetikzlibrary{positioning}
\usepackage[colorlinks=true,linkcolor=blue,citecolor=blue, urlcolor=blue,bookmarks=false]{hyperref}
\usepackage{qcircuit}

\begin{document}
\newtheorem{theorem}{Theorem}[section] 
\newtheorem{lemma}[theorem]{Lemma}

\title{General noise-resilient quantum amplitude estimation}
\date{\today}

\author{Yonglong Ding}
\affiliation{School of Physics, Beijing Institute of Technology, Beijing 100081, China}
\affiliation{Beijing Computational Science Research Center, Beijing 100193, China}
\author{Ruyu Yang}
\email{yangruyu96@gmail.com}
\affiliation{Graduate School of China Academy of Engineering Physics, Beijing 100193, China}

\begin{abstract}
Quantum advantage requires overcoming noise-induced degradation of quantum systems. Conventional methods for reducing noise such as error mitigation face scalability issues in deep circuits. Specifically, noise hampers the extraction of amplitude and observable information from quantum systems. In this work, we present a novel algorithm that enhances the estimation of amplitude and observable under noise. Remarkably, our algorithm exhibits robustness against noise that varies across different depths of the quantum circuits. We assess the accuracy of amplitude and observable using numerical analysis and theoretically analyze the impact of gate-dependent noise on the results. This algorithm is a potential candidate for noise-resilient approaches that have high computational accuracy.
\end{abstract}

\maketitle
Quantum amplitude estimation (QAE) \cite{nielsen2002quantum,cleve1998quantum} is a fundamental technique for quantum computation that has been applied to various fields \cite{ahnefeld2022coherence}, such as quantum chemistry \cite{knill2007optimal,pj2008mohseni}, finance \cite{rebentrost2018quantum}, and machine learning \cite{wiebe2014quantum,kerenidis2019q}. In the field of quantum computing, many efforts have been made to improve QAE algorithms \cite{marvian2022operational,grinko2021iterative}. A promising approach was proposed by Suzuki et al. \cite{suzuki2020amplitude}, which uses a combination of Grover iterations and Maximum Likelihood Estimation (MLE) to replace Quantum Phase Estimation (QPE). Wiebe and Granade \cite{Wiebe2015EfficientBP} developed an efficient Bayesian Phase Estimation. Several novel algorithms, based on Kitaev’s iterative techniques, have also been proposed \cite{kitaev1995quantum,suzuki2020amplitude}. Suzuki et al. \cite{suzuki2020amplitude} and Wie et al. \cite{wie2019simpler} suggested some possible simplifications to QAE, but they did not provide rigorous proof of their correctness. Aaronson et al. \cite{aaronson2020quantum} was the first to rigorously prove that QAE, even without QPE, can achieve a quadratic speedup over classical Monte Carlo simulations \cite{zhou2018achieving}. 

However, performing QAE in noisy environments poses significant challenges that can affect the algorithm's performance. Generally, the depth of the quantum circuit for a desired accuracy $\epsilon_0$ using QAE scales as $O(1/\epsilon_0)$. This means that deeper circuits are often required to achieve higher accuracy. However, in the NISQ era, the resources required to obtain correct results increase exponentially with the depth of quantum circuits~\cite{takagi2023universal,takagi2022fundamental,quek2022exponentially,tsubouchi2023universal}, making it difficult to implement QAE. Some QAE algorithms incorporate an explicit noise model to enhance estimation performance~\cite{brown2020quantum, uno2021modified}. Reducing circuit depth is another research direction~\cite{plekhanov2022variational, giurgica2022low}. However, this often complicates the algorithm.

In this work, we propose noise-resilient quantum amplitude estimation(NRQAE). Under the assumption that the noise is gate-dependent, the calculation results are only affected by the noise of a single layer of quantum circuits and not by the noise in the entire quantum circuit. Specifically, we begin by preparing the quantum states to be estimated into quantum circuits. Next, we execute the Grover operator $G$ \cite{roy2022deterministic,toyama2013quantum,jang2020grover,wong2015grover}, with the trace of $G$ closely related to the overlap between the two quantum states. We prepare several quantum circuits, each differing only in the number of $G$ operators, and determine the quantum amplitude through the algebraic relationship between the different quantum circuits. This approach is motivated by recent work in this field\cite{rall2021faster,plekhanov2022variational,grinko2021iterative,PhysRevLett.68.3121,beals2013efficient,wang2021minimizing,aaronson2020quantum}.  Our work focuses on QAE based on the analysis of different quantum circuits. A detailed analysis yields the total calculation equation, which incorporates noise effects into the algorithm design and makes it more robust\cite{roncallo2023multiqubit,helsen2022general}. We use numerical simulations to show how the accuracy changes with circuit depth under a fixed number of shots and explore the impact of circuit noise on the final results. Moreover, we provide a theoretical explanation of why NRQAE has an advantage over other algorithms.

\textbf{\textit{Algorithm~~~}}
\label{model}
In this section, we explain the details of NRQAE. The main goal is to estimate the amplitude  ${\vert\langle\phi\vert\psi\rangle\vert}^2$ for any two quantum states $\vert\phi\rangle$ and $\vert\psi\rangle$ or the expectation $\langle\psi|O|\psi\rangle$ given the Pauli observable $O$ and state $|\psi\rangle$. The aim of QAE is to achieve the highest precision with the lowest computational expense. Initially, we construct a specialized operator and map it into quantum gates within the quantum circuit.
By manipulating the number of quantum gates and their effects across different circuits operating on the same initial state, we establish an algebraic relationship among multiple quantum circuits to resolve quantum amplitude. This process allows us to attain high-precision amplitude values and enhances noise resilience.

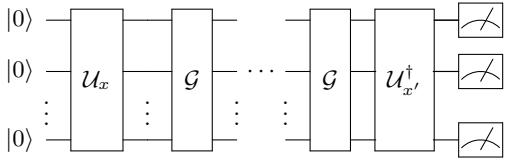
\begin{figure}[t]
	\centerline{
		\Qcircuit @C=1em @R=0.7em {
			\lstick{|0\rangle}  	&  \multigate{3}{\mathcal{U}_{x}}& \qw &\multigate{3}{\mathcal{G}}  & \qw & & & \multigate{3}{\mathcal{G}}&\multigate{3}{\mathcal{U}^{\dagger}_{x^{\prime}} }& \meter  \qw\\
			\lstick{|0\rangle}  	&  \ghost{\mathcal{U}_x}   & \qw &\ghost{\mathcal{G}}    & \qw & \cdots & & \ghost{\mathcal{G}}   & \ghost{\mathcal{U}^{\dagger}_{s^\prime}}  & \meter  \qw  \qw\\
			\vdots                  &   &\vdots    & & \vdots    &  &     \vdots & & \\
			\lstick{|0\rangle}      &  \ghost{\mathcal{U}_x}    & \qw &\ghost{\mathcal{G}}   & \qw & & & \ghost{\mathcal{G}}   &\ghost{\mathcal{U}^{\dagger}_{s^\prime}}  & \meter 
		}
	}
	\caption{\label{fig:1}The circuit structure of control-free phase estimation. The system is first prepared to an initial state by applying a unitary operator $\mathcal{U}_{x}$, with $x=0,1$, corresponds to $|\psi\rangle$ when $x=0$ and to $|\phi\rangle$ when $x=1$, and then a specific number of $\mathcal{G}$ operators are applied to the initial state. Finally, execute $\mathcal{U}^{\dagger}_{s^\prime}$ where $x^{\prime} = 0,1$ and measures the final state on Z basis.}
\end{figure}

In our circuit design, the primary focal point resides in the application of the Grover operator, labeled as $G$, which can be represented as $G = G_0 G_1$. First, we consider the case of estimating the amplitude $|\langle\psi|\phi\rangle|^2$. The states can be encoded into the circuit as $G_0 = 2|\psi\rangle\langle\psi| - I$ and $2|\phi\rangle\langle\phi| - I$, where $I$ is the identity operator.
 The key feature is that we can connect the value ${\vert\langle\psi\vert\phi\rangle\vert}^2$  with the
trace of $G$.
\begin{equation}
	\text{Tr}(G_{0}G_{1}) =4|\langle\psi|\phi\rangle|^2-4 + 2^n. 
\end{equation}
 Because $G$ is unitary, the modulus of its eigenvalue is 1. Therefore, we only need to find the phase $\theta$ of the eigenvalue to find the trace of $G$ and then estimate the amplitude $|\langle\psi|\phi\rangle|^2$.

Subsequently, we delve into an extensive examination of its properties, offering a comprehensive algorithmic framework. Without loss of generality, we can express quantum states $|\psi\rangle$ and $|\phi\rangle$ as
\begin{align}
		\vert\phi\rangle=\cos\mu\vert\alpha\rangle +e^{-i\lambda}\sin\mu\vert\beta\rangle,\\
        \vert\psi\rangle=\cos\nu\vert\alpha\rangle +e^{i\lambda}\sin\nu\vert\beta\rangle,\,\,\,\,
\end{align}
where $|\alpha\rangle$ and $|\beta\rangle$ are orthogonal to each other, and $\mu,\lambda,\nu$ are positional and unimportant parameters. Furthermore, 
We can express $G_0$ and $G_1$ as a matrix using $|\alpha\rangle$ and $|\beta\rangle$ as basis vectors
\begin{align}
   G_0 = 
\begin{bmatrix}
	0 & e^{-i\lambda} \\ e^{i\lambda} & 0
\end{bmatrix}
\begin{bmatrix}
	\cos2\mu & \sin2\mu \\ \sin2\mu & -\cos2\mu
\end{bmatrix}
\begin{bmatrix}
	0 & e^{-i\lambda} \\ e^{i\lambda} & 0
\end{bmatrix},
\label{eq:3} \\
G_1=
\begin{bmatrix}
	0 & e^{i\lambda} \\ e^{-i\lambda} & 0
\end{bmatrix}
\begin{bmatrix}
	\cos2\nu & \sin2\nu \\ \sin2\nu & -\cos2\nu
\end{bmatrix}
\begin{bmatrix}
	0 & e^{i\lambda} \\ e^{-i\lambda} & 0
\end{bmatrix}.
\label{eq:4}
\end{align}

The iterative application of the $G$ operator $n$ times can be expressed as
\begin{align}
&G^n =\nonumber\\ 
&	\begin{bmatrix}
		0 & e^{-i\lambda} \\ e^{i\lambda} & 0
	\end{bmatrix}
	\begin{bmatrix}
		\cos2n(\mu-\nu) & \sin2n(\mu-\nu) \\ \sin2n(\mu-\nu) & -\cos2n(\mu-\nu)
	\end{bmatrix}
	\begin{bmatrix}
		0 & e^{-i\lambda} \\ e^{i\lambda} & 0
	\end{bmatrix}\,\,
\label{eq:5} 	
\end{align}
Where $\lambda$ can be calculated.Ignoring the action of $\lambda$
, then the action of $G$ can be expressed as


\begin{align}
    G^{n}\vert{\alpha}\rangle=\cos\left(2{n}(\mu-\nu)\right)\vert\alpha\rangle+\sin\left(2{n}(\mu-\nu)\right)\vert\beta\rangle,
\label{eq:6} \\
G^{n}\vert{\beta}\rangle=\sin\left(2{n}(\mu-\nu)\right)\vert\alpha\rangle-\cos\left(2{n}(\mu-\nu)\right)\vert\beta\rangle\,\,
\label{eq:7}
\end{align}
\begin{figure}[h]
\includegraphics[width=6.0cm]{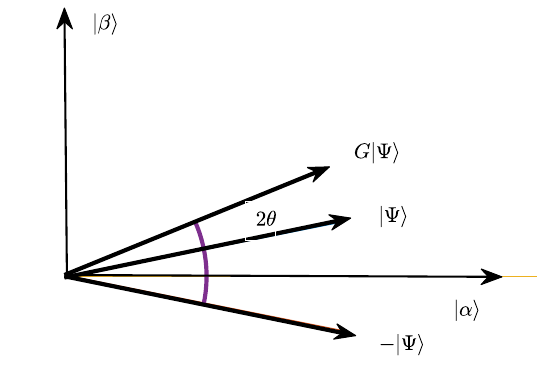}
\caption{\label{fig:2} Under the assumption that $\lambda = 0$, the effect of the $G$ operator on the initial state can be visually expressed as a rotation in two-dimensional space, and the angle of counterclockwise rotation is the same each time. The angle of deflection in the graph is twice the angle between the two states under test. $\vert \Psi \rangle$ represents the entity subject to $G^{\prime}$s operation. }
	\label{fig2}
\end{figure}
A single $G$ operator influences the state by causing a counterclockwise rotation, as depicted in Fig.~\ref{fig:2}. The cumulative effect of multiple $G$ operators on the state is equivalent to a counterclockwise rotation by an angle multiplied by $n$, as described by Eq.~\ref{eq:6} and Eq.~\ref{eq:7}.

Through an iterative scheme, we employ $n$, $2n$, and $3n$ instances of the $G$ operator in three distinct quantum circuits using the same initial state. This process yields measurements $l_{n},l_{2n},l_{3n}$ based on four direct measurements, enabling the calculation of three different measurement outcomes,
\begin{align}
l_{n}:&={\vert \langle \phi \vert G^n \vert \phi\rangle \vert}^2-{\vert \langle \phi \vert G^n \vert \psi\rangle \vert}^2\nonumber\\
&-{\vert \langle \psi \vert G^n \vert \phi\rangle \vert}^2+{\vert \langle \psi \vert G^n \vert \psi\rangle \vert}^2,
\label{post-processing}
\end{align}
calculated from the outcomes of four direct measurements.
By conducting this process with a predefined number of measurement shots, we obtain three different measurement outcomes. 
The measurement results can be represented as 
\begin{align}
 &t_{n}=cp^n \cos(n\theta),\nonumber\\ &t_{2n}=cp^{2n} \cos(2n\theta)\nonumber,\\ &t_{3n}=cp^{3n} \cos(3n\theta),
 \label{Eq:equations}
\end{align}
where the angle $\theta$ is the phase of the eigenvalues of $G$ in $\text{V}$. 
The derivation of Eqs.~\ref{Eq:equations} will be detailed in the next section. Leveraging the measurement data from the three circuits, we define
\begin{equation}
y=\dfrac{t_{n}t_{3n}}{t_{2n}^2}=\dfrac{\cos(n\theta)\cos(3n\theta)}{\cos^2(2n\theta)}.
 \label{Eq:10}
\end{equation}
Combining properties of trigonometric functions, we then formulate a quadratic equation with $\cos(2n\theta)$ as the independent variable and provide the root-finding formula for 
\begin{equation}
\cos(2n\theta)=\dfrac{1\pm\sqrt{1-8(y-1)}}{4(y-1)}.
 \label{Eq:11}
\end{equation}
Solving this equation provides multiple potential solutions of the argument $\theta$, with our process selecting the most precise solution from the previous iteration's results. This iterative refinement progressively enhances the precision of the amplitude results, an aspect clearly illustrated in the accompanying program diagram.

For the case of estimating the expectation of a Pauli observable, the definition of $G$ changes to $G_O = (2|\psi\rangle\langle\psi|-I)O$. 
Correspondingly, the relationship between trace of $G_O$ and expectation $\langle\psi|O|\psi\rangle$ is
\begin{align}
    \text{Tr}(G_O) = 2\langle\psi|O|\psi\rangle.
\end{align}
 Then, we only need to follow the algorithm for estimating amplitude to estimate the eigenvalue phase of $G_O$, and then we can find the trace of $G_O$, and then find the expectation. We summarize the details of NRQAE in TABLE.~\ref{Table}.

\begin{table}
$
\begin{array}{rlr}

\hline\\
&NRQAE(k,G,N_{shots},\rho_1,\rho_2)\\

\hline
1& \theta_u\leftarrow 0 //initialize\,\,angle \\
2 & for \,\,i= 0:1:k   //Set\,\, the\,\, iteration\,\, count\\
3 & \,\,\,\,n\leftarrow 2^{i}//Depth\,\, of\,\, the\,\, circuit\\
4 & \,\,\,\,\theta_l \leftarrow\theta_u \\
5 & \,\,\,\, \Tilde{\rho}=\vert \phi \rangle \langle \phi \vert-\vert \psi \rangle \langle \psi \vert //Initial \,\,state\\
 
6 & \,\,\,\, G^{n}(\rho)  //Quantum \,\,circuit 1\\

7 & \,\,\,\,G^{2n}(\rho)  //Quantum \,\,circuit 2\\

8 & \,\,\,\, G^{3n}(\rho)  //Quantum \,\,circuit 3\\

9 & \,\,\,\, l_{n}=M[G^{n}\rho ] //Measure\,\, each\\
 &\,\,\,\, qubit \,\,in\,\, quantum\,\, circuit\,\, 1 \\

10 & \,\,\,\,l_{2n}=M[G^{2n}\rho] //Measure\,\, each\\
 &\,\,\,\, qubit \,\,in\,\, quantum\,\, circuit\,\, 2 \\
11 &\,\,\,\, l_{3n}=M[G^{3n}\rho] //Measure\,\, each\\
 &\,\,\,\, qubit \,\,in\,\, quantum\,\, circuit\,\, 3 \\
12 &\,\,\,\, \cos(2n\theta)=Function(l_{n},l_{2n},l_{3n})\\
& \,\,\,\,//Solve \,\,the \,\,equation \,\,and \,\,generate \,\,multiple \\
&\,\,\,\,solutions \\
13 &\,\,\,\, \theta_1...\theta_{2n}= \arccos(\cos(2\theta))/2\\
&\,\,\,\,//Figure\,\, out\,\, the\,\, Angle \\
14 & \,\,\,\,\theta_u=min(\theta_1-\theta_l,...,\theta_{2n}-\theta_l)\\
&\,\,\,\,//Choose\,\, the\,\, accurate\,\, solution  \\
\hline
 
\end{array}
$
\caption{\label{Table}The $M$ signifies the measurement of quantum 
circuit amplitudes, whereas $Function$ denotes the 
process outlined by Eq.~\ref{Eq:10} and Eq.~\ref{Eq:11}. }
\end{table}


\textbf{\textit{Error analysis~~}}
In this section, we first discuss the invariant subspace of $G$, and then based on this we discuss the impact of gate-dependent noise on the results.
Specifically, we use the trace of $G$, which is difficult to directly obtain due to its presence in a $2^n$ dimensional Hilbert space. Fortunately, there is a workaround for the case of estimating amplitude by measuring this trace within a subspace
$\text{V}_a=\text{span}(\vert\phi\rangle,\vert\psi\rangle)$. It can be verified that for any state $|s\rangle \in \text{V}$, 
\begin{align}
    G|s\rangle \in \text{V};
\end{align}
and for any state $|s\rangle \in \text{V}_{\bot}$,
\begin{align}
    G|s\rangle = |s\rangle,
    \label{eq:outside_1}
\end{align}
where $\text{V}_{\bot}$ is the complement space of $\text{V}$. Obviously, according to Eq.~\ref{eq:outside_1}, the eigenvalues of $G$ outside subspace $\text{V}$ are all 1. 

For the case of estimating the expectation of the Pauli operator $O$, the corresponding invariant subspace is $\text{V}_O=\text{span}(\vert\psi\rangle,O\vert\psi\rangle)$. The eigenvalues of $G_O$ outside subspace $\text{V}_O$ are all -1. It can be verified that 
for any state $|s\rangle \in \text{V}_{O,\bot}$,
\begin{align}
    G_O|s\rangle = -|s\rangle.
    \label{eq:outside_2}
\end{align}
Comparing the situation of estimating amplitude, we can find that $O\vert\psi\rangle$ actually plays the role of $|\phi\rangle$. 
Thus the following discussion applies to both situations.
Let’s see the property of $G$  in $\text{V}$. 
For simplicity, we expand the state $|\psi\rangle$ as 
\begin{equation}
    |\psi\rangle = a|\phi\rangle + b|\phi_{\bot}\rangle,
\end{equation}
where $|\phi_{\bot}\rangle \in \text{V}$ and $\langle \phi|\phi_{\bot}\rangle = 0$.
The eigenstates of $G$ in the subspace can be represented as:
\begin{equation}
\vert\phi_{+}\rangle=\dfrac{1}{\sqrt{2}}\vert\phi\rangle+\dfrac{i}{\sqrt{2}}\vert\phi_{\bot} \rangle,
\end{equation}	
\begin{equation}
\vert\phi_{-}\rangle=\dfrac{1}{\sqrt{2}}\vert\phi\rangle-\dfrac{i}{\sqrt{2}}\vert\phi_{\bot} \rangle.
\end{equation}	
Thus $|\phi\rangle$ and $|\phi_{\bot}\rangle$ can be expanded by these two eigenstates:
\begin{equation}
\vert\phi\rangle=\dfrac{1}{\sqrt{2}}\vert\phi_{+}\rangle+\dfrac{i}{\sqrt{2}}\vert\phi_{-} \rangle,
\end{equation}	
\begin{equation}
\vert\phi_{\bot}\rangle=\dfrac{1}{\sqrt{2}}\vert\phi_{+}\rangle-\dfrac{i}{\sqrt{2}}\vert\phi_{-} \rangle.
\end{equation}	
And furthermore, $|\psi\rangle$ can be written as
\begin{equation}
	\vert\psi\rangle=\dfrac{a-ib}{\sqrt{2}}\vert\phi_{+}\rangle+\dfrac{a+ib}{\sqrt{2}}\vert\phi_{-} \rangle.
\end{equation}	
The initial state and measurement we need is actually the difference between the density matrices of the two target states
\begin{equation}
   \vert\phi\rangle\langle\phi\vert-\vert\psi\rangle\langle\psi\vert=
   \frac{1+(1-2ab)i}{2}\vert\phi_{+}\rangle\langle\phi_{-}\vert+h.c.
\end{equation}
Note that although $ \Tilde{\rho} = \vert\phi\rangle\langle\phi\vert-\vert\psi\rangle\langle\psi\vert$ is not a quantum state, we can still simulate this process by post-processing as shown by Eq.~\ref{post-processing}.

 All states outside $\text{V}$ are highly degenerate eigenstates. According to matrix perturbation theory, degenerate eigenstates may not remain close to their unperturbed states after perturbation. Setting up such a special initial state and measurement can reduce the impact of noise to a great extent.

\begin{figure}[hptb]
\begin{tikzpicture}

\scope[nodes={inner sep=4,outer sep=4}]
\node[anchor=south east] (a)
  {\includegraphics[width=4cm]{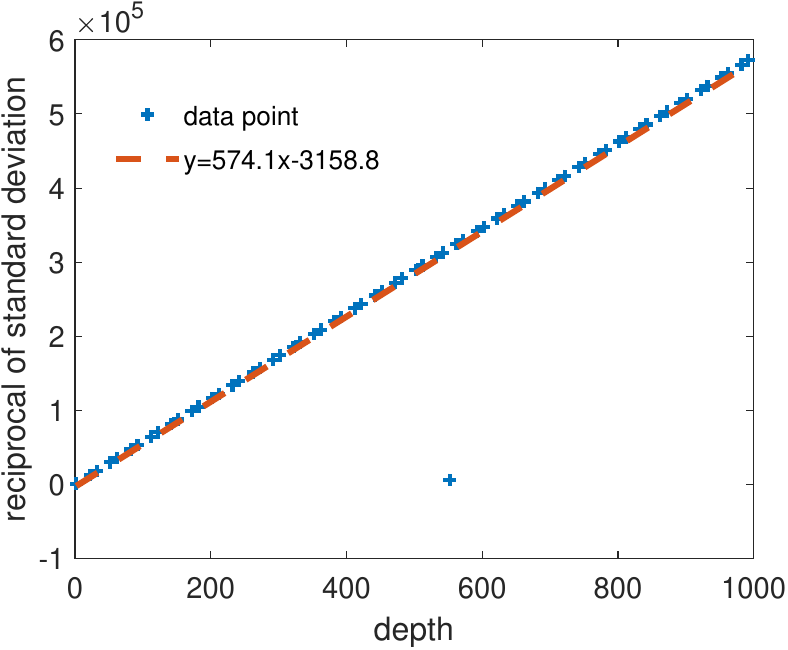}};
\node[anchor=south west] (b)
  {\includegraphics[width=4.0cm]{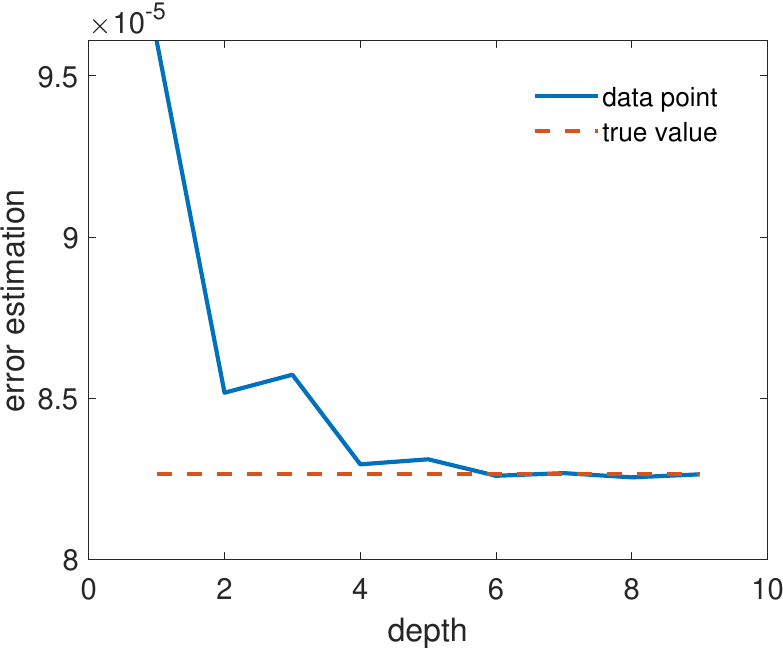}};
\node[anchor=north east] (c)
  {\includegraphics[width=4cm]{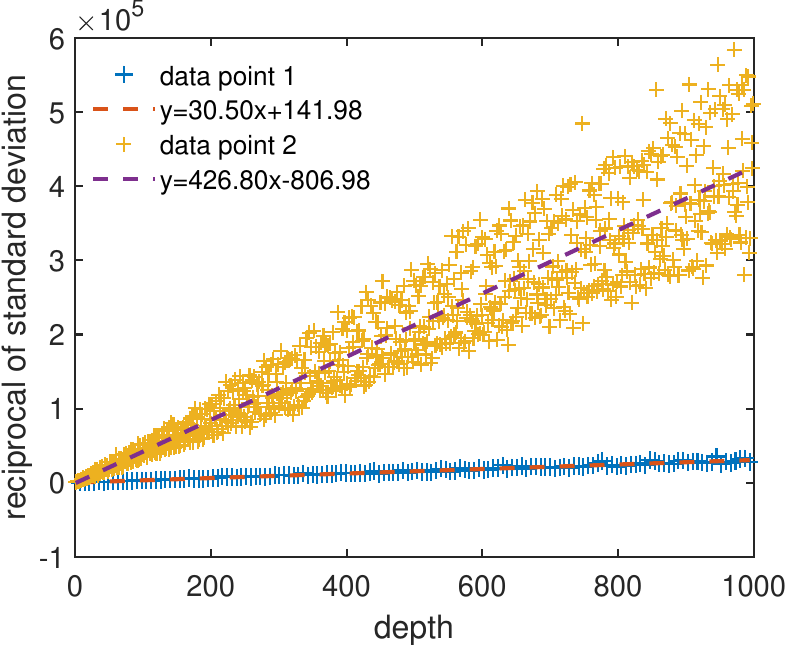}};
\node[anchor=north west] (d)
  {\includegraphics[width=4.0cm]{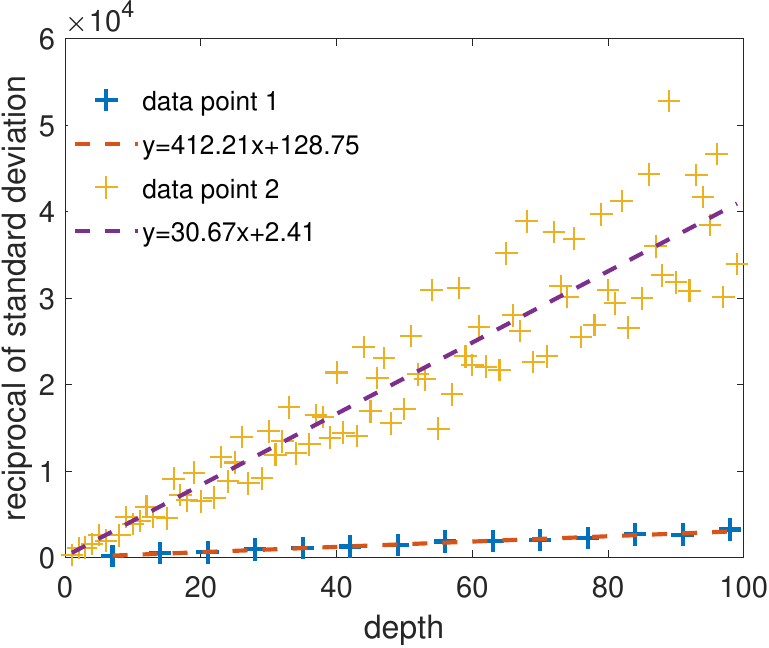}};
\endscope
\foreach \n in {a,b,c,d} {
  \node[anchor=north west] at (\n.north west) {(\n)};
}

\end{tikzpicture}
\caption{\label{fig:6}The simulations in (a)(b)(d) involve the two-qubit scenario, while (c) shows the single-qubit case. (a)The horizontal axis is the depth of the quantum circuit, and the vertical axis is the accuracy of the estimation, With a fixed error of 0.01 added to our measurement outcomes, we observed a gradual reduction in error as the depth increased. Error is the absolute difference between the estimated angle and the ideal angle($\pi/3$). Accuracy is defined as the inverse of error. (b)The accuracy increases with increasing depth. The term "true value" denotes the perturbative argument $\theta$ in Eq.~\ref{eq: equation_theta}(ideally $\pi/5$).  (c)The vertical axis represents the standard deviation of the results(ideally $\pi/7$), which decreases as the depth increases.  (d)The vertical axis represents the standard deviation of the results(ideally $\pi/7$), which decreases as the depth increases. }
	\label{fig6}
\end{figure}

To analyze noisy quantum circuits, we use the vectorization of the density matrix and the matrix representation of the quantum channel. A straightforward method is to vectorize the density matrix as:
\begin{equation}
\rho=\Sigma_{i,j}a_{i,j}\vert\psi_{i}\rangle\langle\psi_{j}\vert\rightarrow\vert\rho\rangle\rangle=\Sigma_{i,j}a_{i,j}\vert\psi_{i}\rangle\otimes\vert\psi_{j}\rangle.
\label{vec}
\end{equation}	
Note that this is not the only way to vectorize a density matrix (another common method in quantum information theory is the Pauli transfer matrix representation). Using Eq.~\ref{vec}, we can express a general map applied to density matrix $\rho$ as a matrix,
\begin{equation}
	 A\rho B\rightarrow A\otimes B^{\dagger}\vert\rho\rangle\rangle.
  \label{vec: gate}
\end{equation}	
The inner product of two vectors is defined as
\begin{equation}	\langle\langle\sigma\vert\rho\rangle\rangle=Tr(\sigma^{\dagger}\rho).
\end{equation}	
Following Eq.~\ref{vec: gate}, we can write the operator $G$ in the invariant subspace $\text{V}$ as
\begin{equation}
	M_{G}=
	\begin{bmatrix}
		1 & 0 & 0 & 0 \\ 0 & e^{2i\theta} & 0 & 0 \\  0 & 0 & e^{-2i\theta} & 0 \\  0 & 0 & 0 & 1 
	\end{bmatrix}.
\end{equation}
The eigenvectors of $M_{G_{0}G_{1}}$ are $\vert\phi_{+}\rangle\otimes\vert\phi_{+}\rangle$,$\vert\phi_{+}\rangle\otimes\vert\phi_{-}\rangle$,$\vert\phi_{-}\rangle\otimes\vert\phi_{+}\rangle$
and
$\vert\phi_{-}\rangle\otimes\vert\phi_{-}\rangle$,
respectively.
Outside the subspace $\text{V}$, the eigenvectors have these forms: 
$\vert\phi_{+}\rangle\otimes\vert\Phi\rangle$,$\vert\phi_{-}\rangle\otimes\vert\Phi\rangle$,$\vert\Phi\rangle\otimes\vert\phi_{+}\rangle$,$\vert\Phi\rangle\otimes\vert\phi_{-}\rangle$, and $|\Phi_{i}\rangle\otimes |\Phi_{j}\rangle$ where $|\Phi\rangle$, $|\Phi_{i}\rangle$ and $|\Phi_{j}\rangle $ are arbitrary states outside $\text{V}$.
 Their eigenvalues are  $e^{\pm i\theta}$ and $\pm 1$. In general, these eigenvectors
are highly degenerate. When the quantum channel $G$ is perturbed, the perturbed degenerate eigenvectors become a randomly linear combination of the degenerate eigenvectors. On the other hand, the non-degenerate eigenvectors change slightly. Therefore, we need to discard the cross-term in the initial state and measurement.

 Without loss of generality, we assume the quantum noise is gate-dependent and model the noisy quantum channel as
 $NM_{G}$.
 The perturbed non-degenerate eigenvalues are 
 $\lambda_{1}$ and $\lambda_{2}$, 
 corresponding to the eigenvectors $\vert\rho_{1}\rangle\rangle$,
 $\vert\rho_{2}\rangle\rangle$,  
 respectively. The difference between the perturbated eigenvectors and the ideal eigenvectors are
  $\vert\bigtriangleup_{1}\rangle\rangle=\vert\rho_{1}\rangle\rangle-\vert\phi_{+}\rangle\otimes\vert\phi_{-}\rangle$ 
 and $\vert\bigtriangleup_{2}\rangle\rangle=\vert\rho_{2}\rangle\rangle-\vert\phi_{-}\rangle\otimes\vert\phi_{+}\rangle$. Remark that both $\vert\rho_{1}\rangle\rangle$ and $\vert\rho_{2}\rangle\rangle$  are only relevant to
the noise in one layer $G$. Using $\epsilon$ to denote $\|NM_G - M_G\|$, where $\|\cdot\|$ is the matrix norm, the initial state 
$\vert\Tilde{\rho}\rangle\rangle=\vert\phi_{+}\rangle\otimes\vert\phi_{+}\rangle-\vert\phi_{-}\rangle\otimes\vert\phi_{-}\rangle$
can be
expressed as
\begin{equation}
\vert\Tilde{\rho}\rangle\rangle=c_{1}\vert\rho_{1}\rangle\rangle-c_{2}\vert\rho_{2}\rangle\rangle+O(\epsilon),
\label{eq:21}
\end{equation}  
  where both $\vert c_{1}-c\vert$ and 
  $\vert c_{2}-c^{\ast}\vert$ 
  are in order of $O(\epsilon)$ and $c= \left(1+(1-2ab)i\right)/2$. After executing the quantum channel $G$ $n$ times in the circuit, the measurement results can be
  written as
 \begin{equation}
\langle\langle\Tilde{\rho}\vert(NM_{G})^{n}\vert\Tilde{\rho}\rangle\rangle=\vert c_{1}\vert^{2}\lambda_{1}^n+\vert c_{2}\vert^{2}\lambda_{2}^n+O(\epsilon)
 \end{equation}  
In general, the two eigenvalues are conjugate, i,e, $\lambda_{1}=\lambda_{2}^\ast$. The next step is to solve the equations
\begin{equation}
    t_{n}\approx c(\lambda_{1}^n+\lambda_{2}^n)=cp^n \cos(n\theta).
    \label{eq: equation_theta}
\end{equation} 
where we have express the eigenvalues as $\lambda_{1}=pe^{i\theta}$ and $\lambda_{2}=pe^{-i\theta}$. The error arises
from the difference between $c_{1}$ and $c_{2}$, and the error term $O(\epsilon)$ in Eq.~\ref{eq:21}. Both
these errors are in order of $O(\epsilon)$. Please refer to the Appendix D for details. Therefore, the error of the measured value is in the order of $O(\epsilon)$, and the error of the final result is also in the order of $O(\epsilon)$.

\textbf{\textit{Results and discussion~~~}}

Previous sections have elaborated the steps of NRQAE and provided theoretical proof of its effectiveness. Our focus now turns to using numerical simulations to demonstrate and validate the efficiency of our approach. 

First, we test the calculation accuracy of the algorithm and verify that the calculation accuracy increases linearly with the depth of the quantum circuit, as shown in Fig.~\ref{fig:6}(a).
Moreover, Fig.~\ref{fig:6}(b) shows the fast convergence of computational results to the phase of perturbative eigenvalues with increasing circuit depth, further confirming the NRQAE’s high accuracy and efficiency.

\begin{figure}[t]
\begin{tikzpicture}

\scope[nodes={inner sep=4,outer sep=4}]
\node[anchor=south east] (a)
  {\includegraphics[width=4cm]{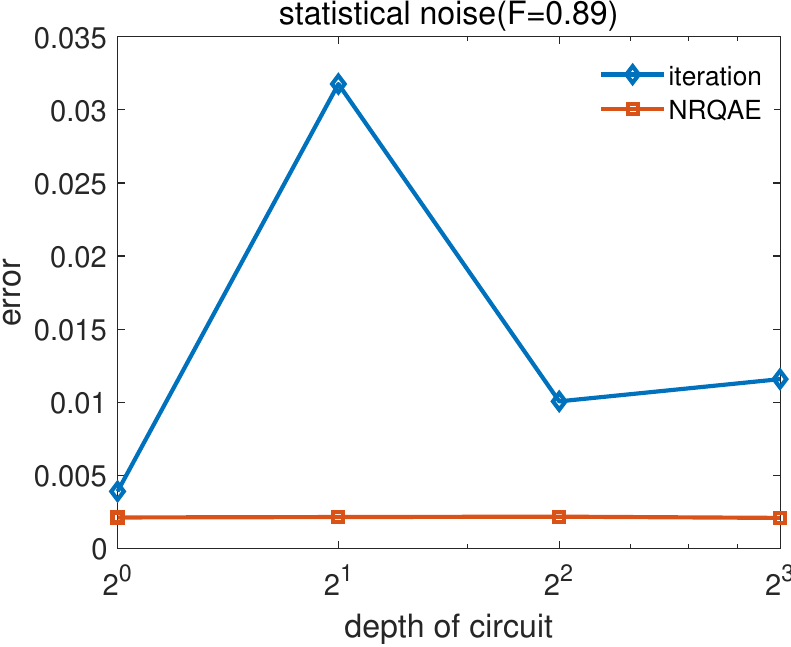}};
\node[anchor=south west] (b)
  {\includegraphics[width=4cm]{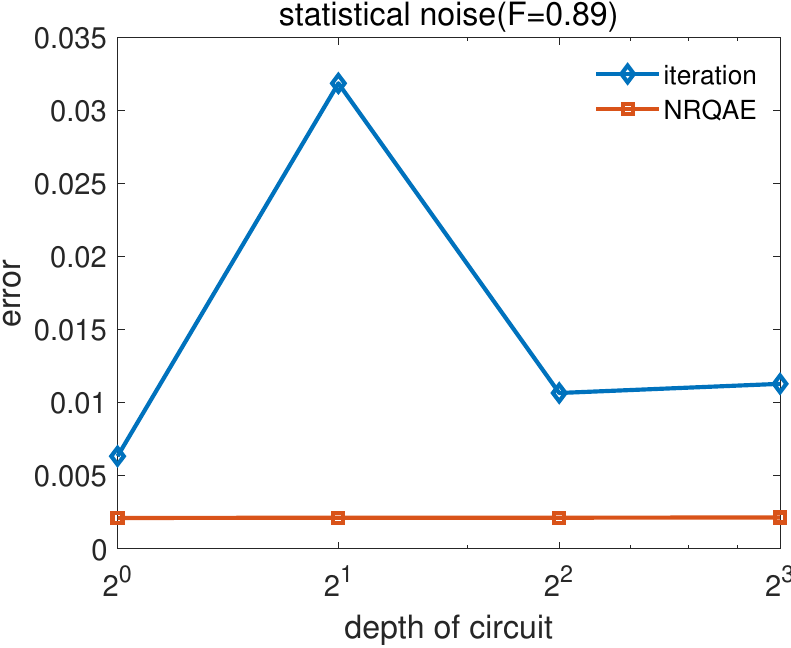}};
\node[anchor=north east] (c)
  {\includegraphics[width=4cm]{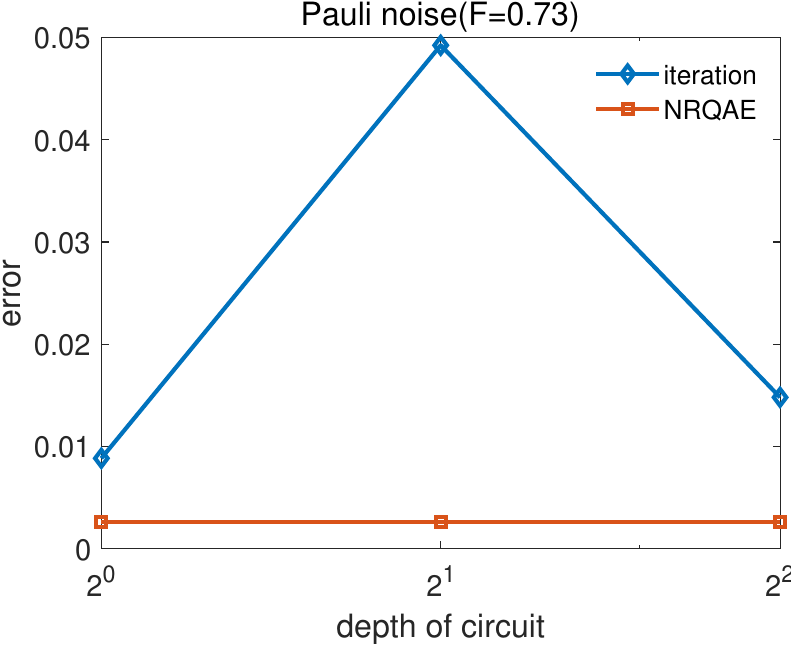}};
\node[anchor=north west] (d)
  {\includegraphics[width=4cm]{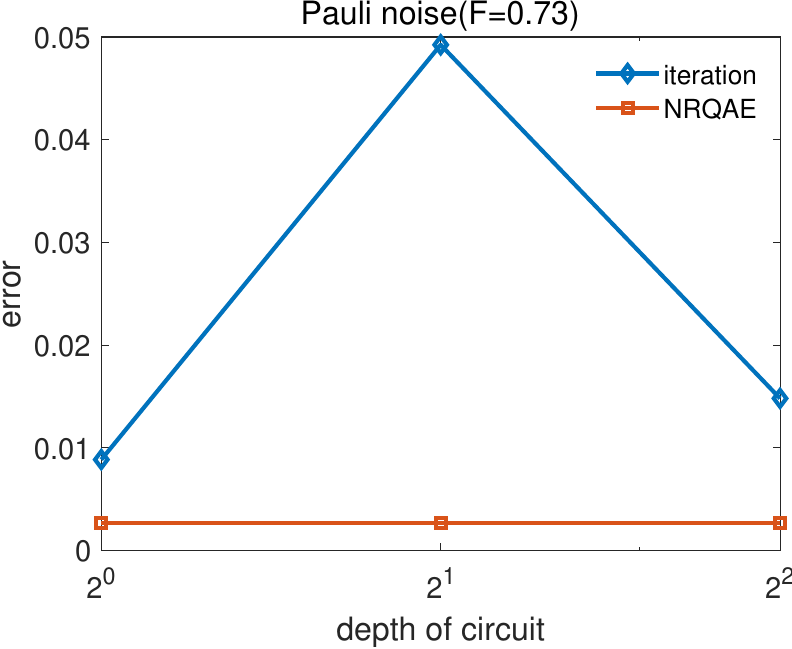}};
\endscope
\foreach \n in {a,b,c,d} {
  \node[anchor=north west] at (\n.north west) {(\n)};
}
\end{tikzpicture}
\caption{\label{fig:44}We simulated a single-qubit system with statistical noise using PTM representation. The blue line shows the result of iterative amplitude estimation, and the orange line shows the result of NRQAE. (a)(b) have statistical noise with a fidelity of 0.89. (c)(d) have Pauli noise with a fidelity of 0.73. (a)(c) are the observation type and the ideal expectation value is 0.1, while (b)(d) are the amplitude type and the ideal amplitude is 0.9. }
	\label{fig4}
\end{figure}

Subsequent simulations, initiated in a noise-free setting, consider single-qubit and two-qubit scenarios, exhibiting a consistent increase in measurement accuracy as the circuit depth expands, shown in Fig.~\ref{fig:6}(c) and (d).
Further analyses scrutinize the quantum circuit's response to diverse noise types introduced into each $G$ operator, encompassing stochastic, amplitude damping, Pauli, coherent, and depolarizing noise. A comprehensive comparative study against established QAE algorithms~\cite{grinko2021iterative} reveals our algorithm's significant advantages in noisy environments, particularly against statistical and Pauli noise, evident in Fig.~\ref{fig:44}.
Appendix Section A expands on NRQAE's robust resilience across varied noise types, demonstrating its applicability in noisy quantum applications. This resilience extends to both amplitude and observation scenarios, underscoring the algorithm's versatile performance in challenging noise-laden settings.

\textbf{\textit{Conclusion~~~}}
\label{conclusion}
In this paper, we present an innovative algorithm aimed at resolving the challenge of estimating the expectation value between two quantum states. We initially transform the quantum state measurement problem by integrating it into the Grover operator framework, translating it subsequently into a sequence of quantum gates within a quantum circuit.

Our extensive analysis explores the behavior of the $G$ operator on quantum states in noisy environments, guiding the selection of an initial "state" and formulating a fitting measurement strategy. Aligned with our design, our quantum circuit exhibits substantial noise resilience without compromising computational accuracy. Supported by rigorous calculations, we offer a theoretical framework explaining the principle of noise resistance, showcasing the substantial enhancements NRQAE provides in accuracy and noise resilience over prevalent QAE methods.

Our method offers distinct advantages. It excels in computational efficiency by dynamically regulating the iteration circuits as their depth increases exponentially. Leveraging the Grover operator, NRQAE significantly bolsters measurement precision. Moreover, it demonstrates robustness in the presence of noise by integrating noise-mitigation techniques within the quantum circuits.

In summary, our novel algorithm provides an efficient and innovative approach to estimate the expectation value between two quantum states. Its potential applications span quantum information processing domains like quantum simulation and machine learning. This noise mitigation approach holds promise for tasks involving the repetitive application of a single operator in quantum circuits, such as Hamiltonian simulation—areas ripe for future exploration.

\textbf{\textit{Acknowledgments---}}
\label{acknowledgments}
This work is supported by National Natural Science Foundation of China (Grant No. 12225507, 12088101) and NSAF (Grant No. U1930403).

\nocite{*}

\bibliography{new}

\onecolumngrid
\newpage 
\newcounter{equationSM}
\newcounter{figureSM}
\newcounter{tableSM}
\newcounter{sectionSM}
\stepcounter{equationSM}
\setcounter{equation}{0}
\setcounter{figure}{0}
\setcounter{table}{0}
\setcounter{section}{0}
\makeatletter
\renewcommand{\theequation}{\textsc{sm}-\arabic{equation}}
\renewcommand{\thefigure}{\textsc{sm}-\arabic{figure}}
\renewcommand{\thetable}{\textsc{sm}-\arabic{table}}
\renewcommand{\thesection}{\textsc{sm}-\arabic{section}}

\begin{center}
  {\large{\bf Supplemental Material for\\
  ``General noise-resilient quantum amplitude estimation''}}
\end{center}
\begin{appendices}
\appendix

\section{A: Comparison of noise models}\label{sm-A}
In this section, we compare NRQAE with the iterative algorithm in five different noise scenarios, showing a clear advantage for our approach. NRQAE demonstrates robust noise mitigation for various noise types.
\begin{figure}[hptb]
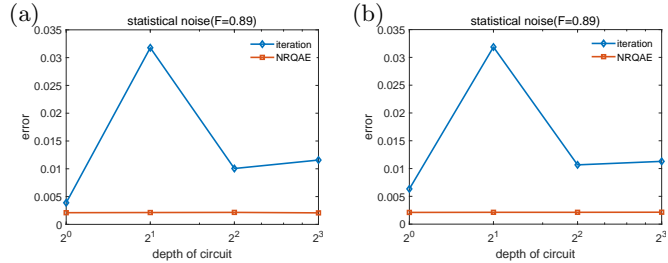

\begin{tikzpicture}
\scope[nodes={inner sep=4,outer sep=4}]
\node[anchor=south east] (a)
  {\includegraphics[width=4cm]{statisticalob.pdf}};
\node[anchor=south west] (b)
  {\includegraphics[width=4cm]{statisticalam.pdf}};
\endscope
\foreach \n in {a,b} {
  \node[anchor=north west] at (\n.north west) {(\n)};
}
\end{tikzpicture}
\caption{\label{fig:11} Statistical noise. The blue line shows the result of iterative amplitude estimation, and the orange line shows the result of NRQAE, with a fidelity of 0.89. (a) is the observation type, while (b) is the amplitude type.}
\label{fig11}
\end{figure}

Quantum stochastic noise can arise from the interactions between quantum systems and the environment. These fluctuations, resulting from uncontrollable factors such as thermal effects and electromagnetic field interactions, can cause decoherence in quantum operations, posing challenges to preserving the desired quantum state.
Fig.~\ref{fig:11} shows the iterative algorithm with the blue line and our approach with the yellow line, highlighting our method's noise mitigation advantage.

\begin{figure}[hptb]
\begin{tikzpicture}

\scope[nodes={inner sep=4,outer sep=4}]
\node[anchor=south east] (a)
  {\includegraphics[width=4cm]{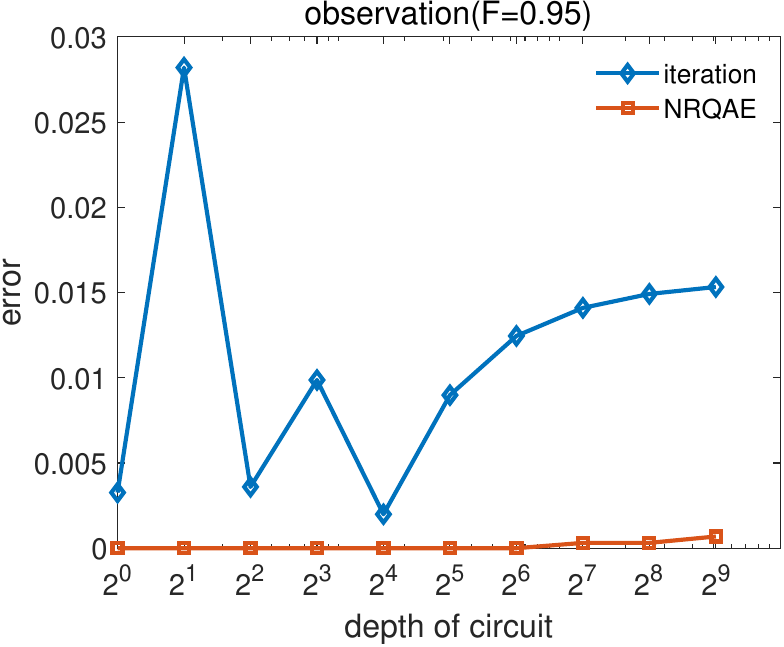}};
\node[anchor=south west] (b)
  {\includegraphics[width=4cm]{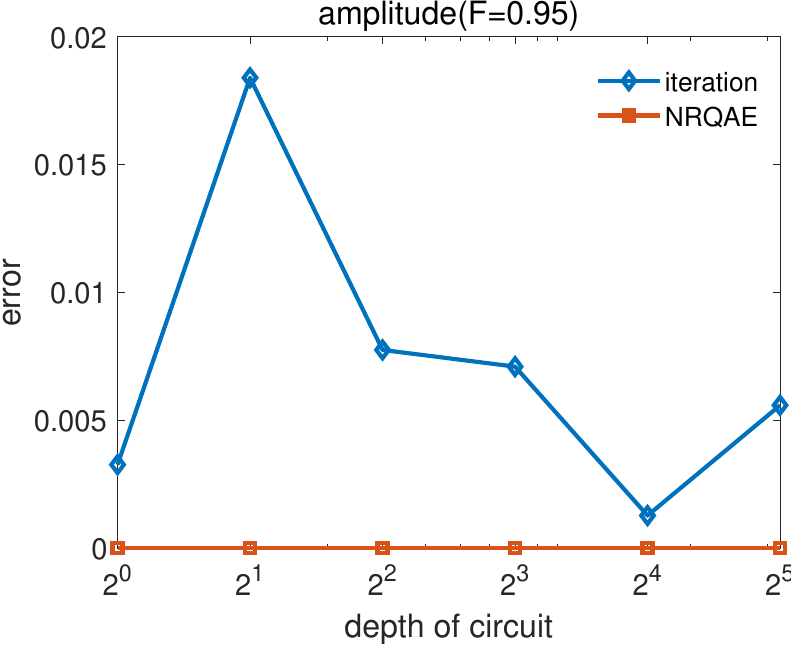}};
\endscope
\foreach \n in {a,b} {
  \node[anchor=north west] at (\n.north west) {(\n)};
}

\end{tikzpicture}
\caption{\label{fig:12}Amplitude damping noise. The blue line represents the result of iterative amplitude estimation, and the orange line represents the result of NRQAE, with an Amplitude-damping (a) is the observation type, while (b) is the amplitude type. }
	\label{fig12}
\end{figure}

Quantum amplitude damping noise specifically impacts quantum systems by degrading their information-carrying capacity and coherence. This noise leads to the gradual reduction of quantum states' amplitudes, often toward a specific state within the system, causing information loss and reduced coherence. Addressing amplitude-damping noise is critical in quantum technologies for maintaining the precision and reliability of quantum operations.
In our program, we model the amplitude damping noise as $N_{noise} = I \times 0.9 + M_{0,0} \times 0.1 + M_{0,1} 0.1$, where $M_{0,0}$ and $M_{0,1}$ are PTM of $|0\rangle\langle 0|$ and $|0\rangle\langle 1|$, respectively. We then compare the iterative algorithm and our approach under these noise conditions, as shown in Fig.~\ref{fig:12}. The results of the iterative algorithm (blue line) and our algorithm (yellow line) are illustrated. Our method exhibits a significant advantage in handling this noise scenario.

\begin{figure}[hptb]
\begin{tikzpicture}

\scope[nodes={inner sep=4,outer sep=4}]
\node[anchor=south east] (a)
  {\includegraphics[width=4cm]{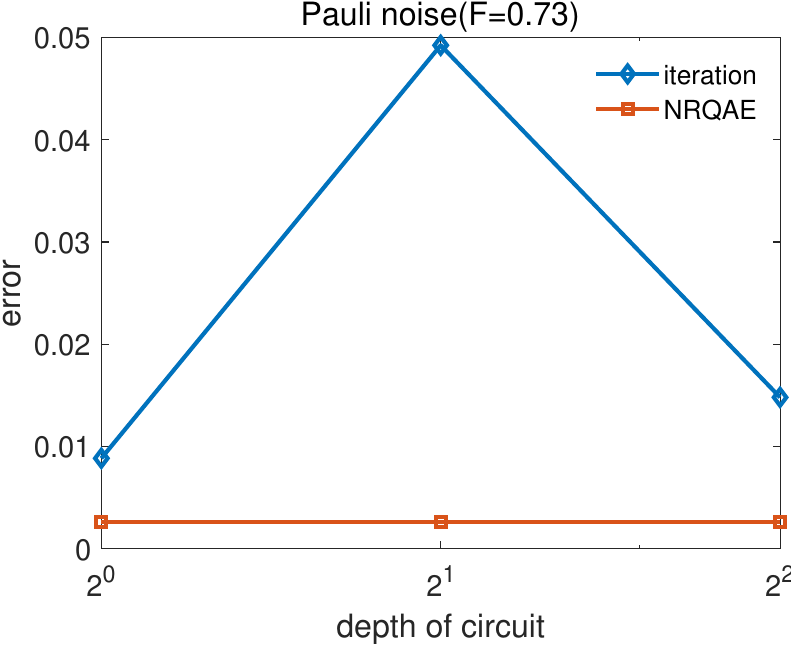}};
\node[anchor=south west] (b)
  {\includegraphics[width=4cm]{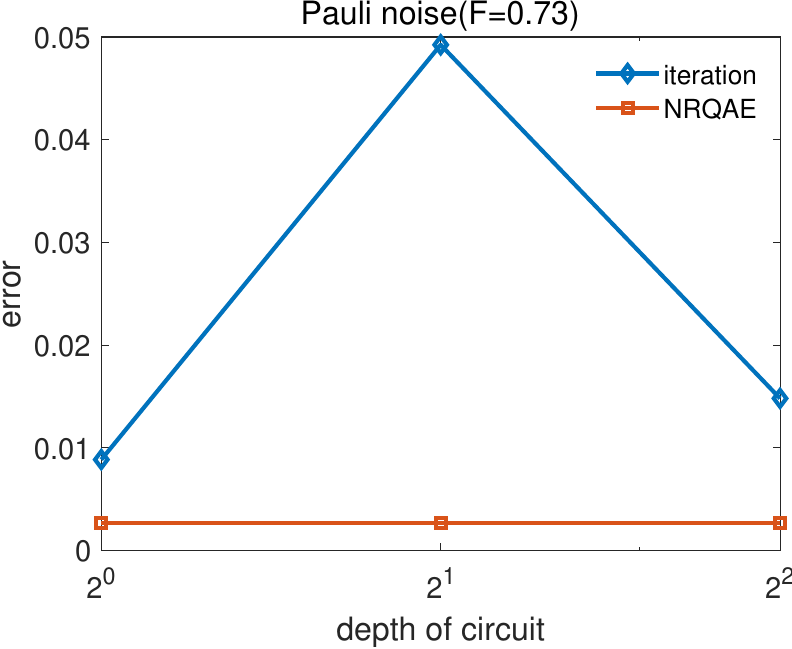}};
\endscope
\foreach \n in {a,b} {
  \node[anchor=north west] at (\n.north west) {(\n)};
}

\end{tikzpicture}
\caption{\label{fig:13}Pauli noise. The blue line shows the result of iterative amplitude estimation, and the orange line shows the result of NRQAE, with a fidelity of 0.80. (a) is the observation type, while (b) is the amplitude type. }
	\label{fig13}
\end{figure}

Pauli noise is very significant in quantum information theory and is a common noise model.
It is characterized by errors represented by Pauli matrices, that cause unintended flips or rotations of quantum states, potentially disrupting quantum computations.
In our program, we model the Pauli noise as $N_{noise} = I \times 0.6 + (\sigma_{x}\times 0.5 +\sigma_{z}\times 1.5)\times 0.2$, where both $\sigma_x$ and $\sigma_z$ are in PTM formalism, enabling a comparison between the iterative algorithm and our approach under this noise scenario. The graph shows the iterative algorithm with the blue line and our algorithm with the yellow line. Our approach exhibits a clear advantage under this noise condition, as shown in Fig.~\ref{fig:13}.

\begin{figure}[hptb]
\begin{tikzpicture}

\scope[nodes={inner sep=4,outer sep=4}]
\node[anchor=south east] (a)
  {\includegraphics[width=4cm]{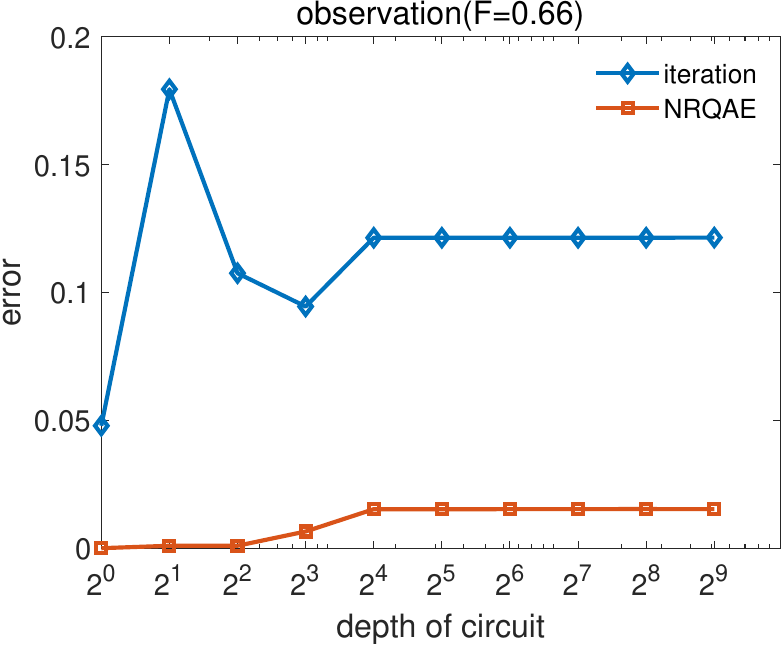}};
\node[anchor=south west] (b)
  {\includegraphics[width=4cm]{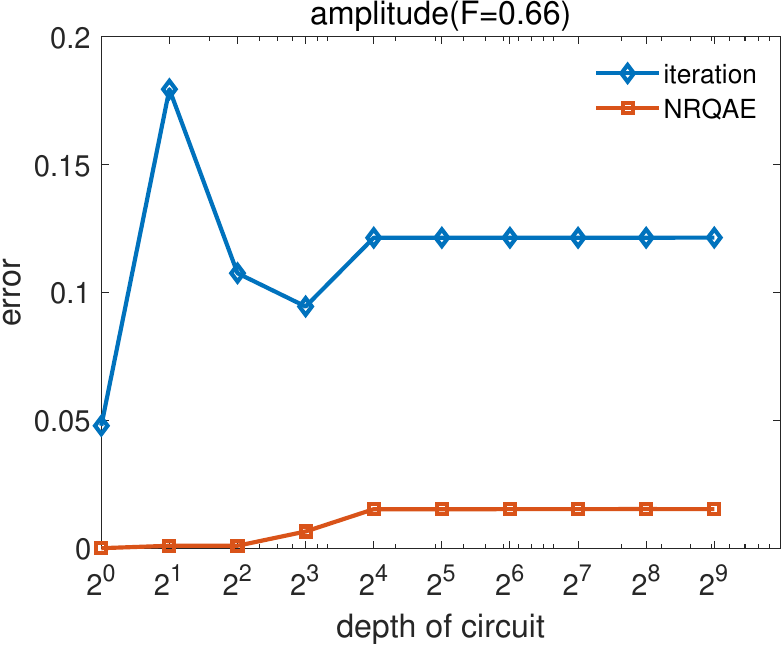}};
\endscope
\foreach \n in {a,b} {
  \node[anchor=north west] at (\n.north west) {(\n)};
}

\end{tikzpicture}
\caption{\label{fig:14}Coherent noise. The blue line shows the result of iterative amplitude estimation, and the orange line shows the result of NRQAE, with a fidelity of 0.99. (a) is the observation type, while (b) is the amplitude type. }
	\label{fig14}
\end{figure}

Quantum coherent noise encompasses various noise types that induce systematic, coherent errors in quantum systems. Unlike stochastic noise, which is random and independent, coherent noise affects the evolution of quantum states in a structured and often deterministic way. External perturbations, control imperfections, or interactions with the environment can cause coherent noise, resulting in phase errors or changes in qubit coherence. These perturbations affect the accuracy of quantum operations and computations. Effectively managing coherent noise is essential in quantum information processing to minimize its effects and preserve the integrity of quantum states.
In our program, we model the coherent noise as $U_{noise} = e^{i\sigma_{x}\delta_{t}}$, $N = U_{noise}\otimes U_{noise}^{\dagger}$, where $\sigma_x$ denotes the Pauli-X operator. enabling a comparison between the iterative algorithm and our approach under this noise scenario. The graph illustrates the iterative algorithm with the blue line and our algorithm with the yellow line. Our approach demonstrates a notable advantage under this noise condition, as shown in Fig.~\ref{fig:14}.

\begin{figure}[hptb]
\begin{tikzpicture}

\scope[nodes={inner sep=4,outer sep=4}]
\node[anchor=south east] (a)
  {\includegraphics[width=4cm]{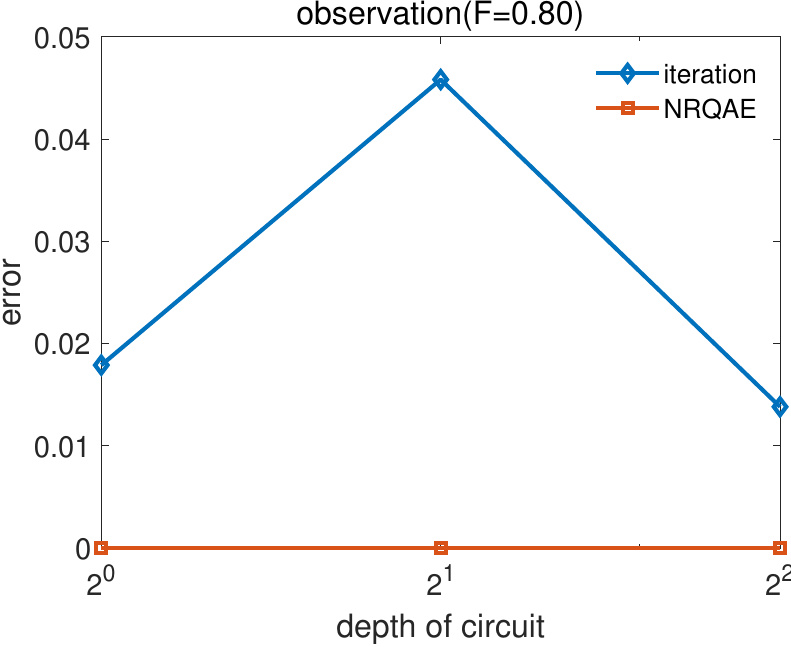}};
\node[anchor=south west] (b)
  {\includegraphics[width=4cm]{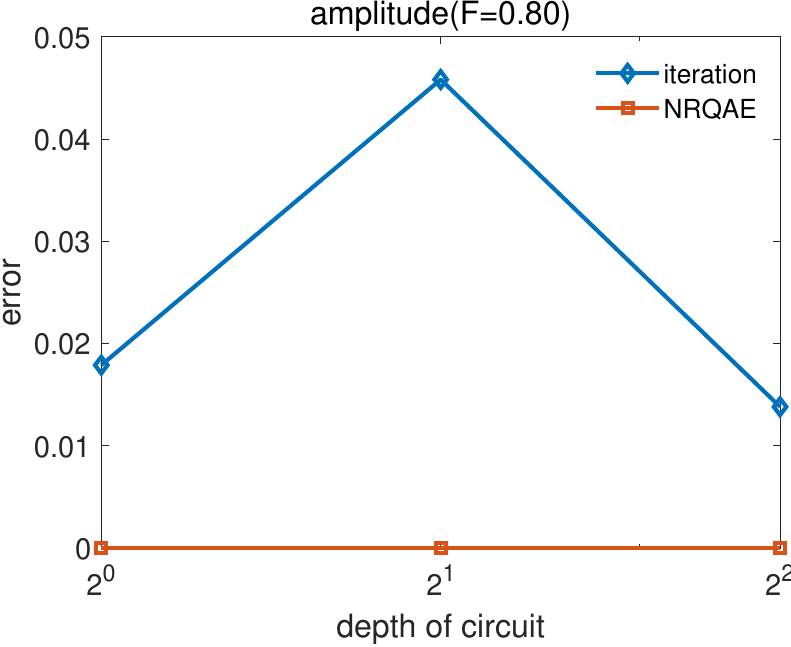}};
\endscope
\foreach \n in {a,b} {
  \node[anchor=north west] at (\n.north west) {(\n)};
}

\end{tikzpicture}
\caption{\label{fig:15} Depolarising noise. The blue line shows the result of iterative amplitude estimation, and the orange line shows the result of NRQAE, with a fidelity of 0.73. (a) is the observation type, while (b) is the amplitude type. }
	\label{fig15}
\end{figure}

Depolarizing noise poses a common challenge in quantum computing and information processing, causing a decline in the fidelity of quantum states and operations. This degradation occurs as quantum systems interact with their environment, resulting in the loss of quantum information and a reduction in state purity.
In our program, we model the noise as $N_{noise} = I \times 0.7 + (\sigma_{x} +\sigma_{z} + \sigma_{y})\times 0.1$, allowing a comparison between the iterative algorithm and our approach under this noise scenario. Here $\sigma_{x}$, $\sigma_y$, $\sigma_y$ are in PTM formalism. we conduct a numerical analysis to assess the algorithm’s reliability, as shown in Fig.~\ref{fig:15}. The Figure illustrates the iterative algorithm with the blue line and our approach with the yellow line, showcasing our method’s superior performance in the face of this noise condition.




\section{B: confidence interval}\label{sm-B}
The previous discussion introduced a specific solution, but its correctness and achievement of ideal precision remain uncertain. To statistically evaluate the algorithm's confidence, we utilized the binomial distribution $P(H(m)\le r)=\sum_{j=0}^r C^{j}_{m}p^{j}(1-p)^{m-j}$ to model the probability of a quantum state yielding 1 $(p)$ or 0 $(1-p)$ upon measurement $H(m)$. Hoeffding's inequality $P(|H(m)-pm|> \varepsilon m)\le 2e^{-2\varepsilon^2 m}$ was applied to calculate the confidence interval, denoted as $\varepsilon$ , providing a high-probability narrow interval for a significant number of measurement points $(m)$. The amplified measured angle by a factor of $n$ corresponds to a similar amplification in the confidence interval within the first quantum circuit.

However, when the number of $G$ operators acting on the initial state is $n$, the deflection angle may surpass $\pi$, resulting in multiple solutions. The proposed solution indicates that if the angle's interval is $\varepsilon$, then $n \varepsilon$ equals $\pi$, accommodating precisely two solutions from the second quantum circuit. A further amplification by a factor of $2n$ contains four solutions in the third circuit. Calculations reveal a high probability of selecting the true solution through the algorithm.
Initially, we verified the quantum circuit's efficacy in multiplying the deflection angle caused by the $G$ operator. We further discussed the variance in computational accuracy with changes in the quantum circuit depth. Finally, we explored the algorithm's stability, closely tied to its confidence level.
Further verification is required to ascertain if the constructed quantum circuit multiplies angles as indicated in Eq.~\ref{eq:5}. A test involving randomly selected $G$ operators, altering their quantity in the initial quantum circuit, confirmed agreement with the results of Eq.~\ref{eq:5}.

Our goal is heightened accuracy with fewer shots, achieved through an increase in the quantum circuit's depth. Analyzing the quantum circuit's accuracy as the depth increased from 20 to 200 resulted in a precision assessment when compared to accurate solutions as shown in Fig.~\ref{fig:17}(c). The section on confidence intervals discussed the probability of the algorithm selecting a successful solution, intrinsically linked to its stability. The main text presented numerical validations, affirming the algorithm's consistent identification of high-precision target solutions with an exceptionally high success rate. 

\section{C: numerical calculation}\label{sm-c}
To explore whether NRQAE has a wide range of applications, we conduct more numerical simulations. It also shows better calculation results.

\begin{figure}[hptb]
	\includegraphics[width=7.0cm]{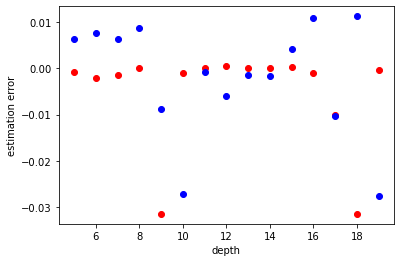}
	\caption{\label{fig:10}The picture shows that the red points are after NRQAE is used, while the blue points are the points without algorithm optimization. Except for a few points, most of the red points have higher precision.}
	\label{fig10}
\end{figure}

To efficiently determine the amplitude, we initially overlook the impact of $\lambda$ when its value is calculable. Referring to Eq.~\ref{eq:5}, when a $G$ operator acts on the initial state, the initial state's deflection angle is $2(\mu-\nu)$. As the count of $G$ operators on the initial states reaches $n$, the initial state's deflection angle becomes $2n(\mu-\nu)$.

In this simulation, we directly measure the final states' amplitudes. However, in quantum computing, each calculation's outcome fluctuates within a small range, mainly influenced by the number of shot points in our algorithm. Consequently, varying deflection angles produce similar fluctuation ranges in measurements with identical shot numbers, ensuring equivalent measurement accuracy for both angles. Given the magnification of the angle by a factor of $n$, the resulting accuracy also scales proportionately.
Notably, without limiting the size of $n$, the value $2n(\mu-\nu)$ might exceed $\pi$, resulting in multiple solutions.

To address this, we construct three quantum circuits. The first employs a single $G$ operator on the initial state, while the second and third circuits use numerous $G$ operators on initial states of $n$ and $2n$, respectively. The amplification of the value interval for $n$ times aims to be $\pi$, allowing the identification of two and four solutions in the second and third quantum circuits, respectively. The real solution among these multiples can be determined based on the proximity between solutions. The one with the smallest distance is selected as the real solution.

\begin{figure}[hptb]
\begin{tikzpicture}

\scope[nodes={inner sep=4,outer sep=4}]
\node[anchor=south east] (a)
  {\includegraphics[width=4cm]{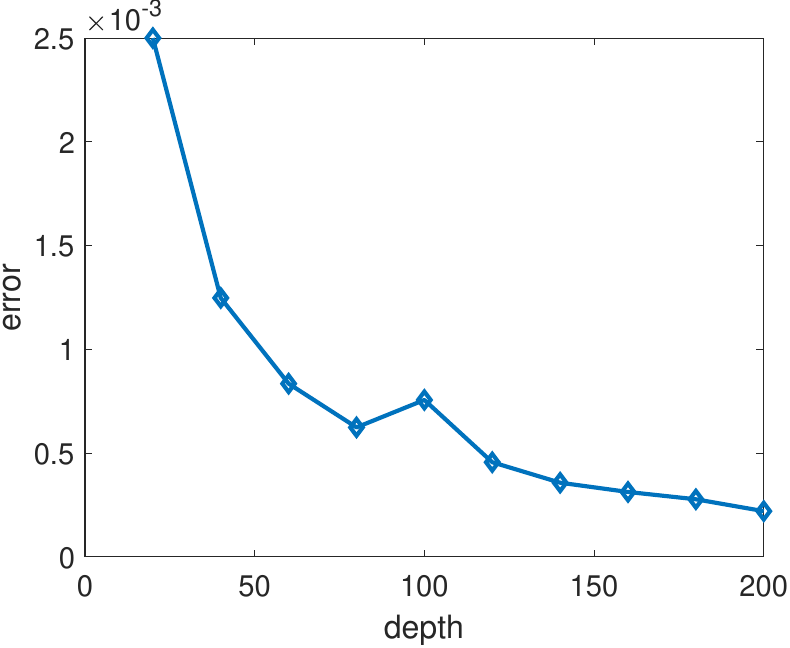}};
\node[anchor=south west] (b)
  {\includegraphics[width=4cm]{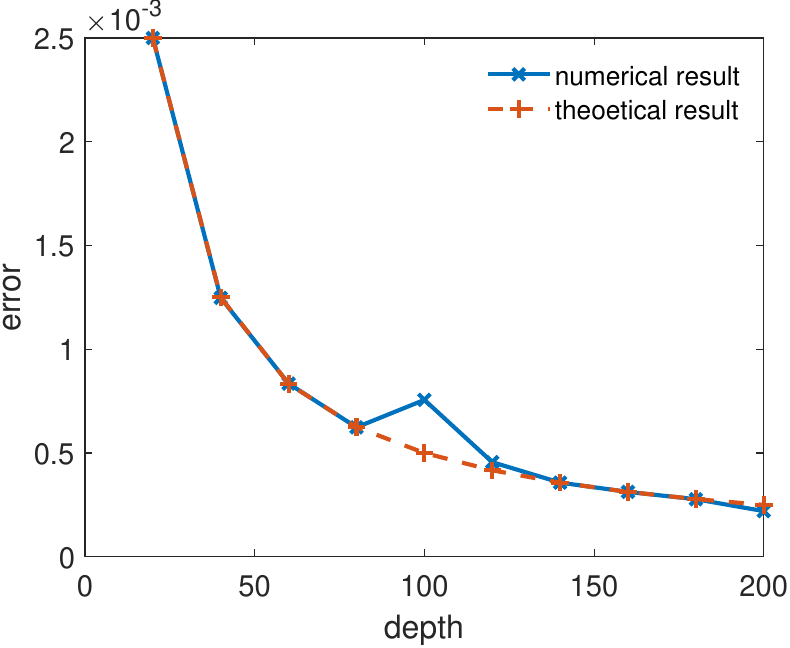}};
\node[anchor=north east] (c)
  {\includegraphics[width=4cm]{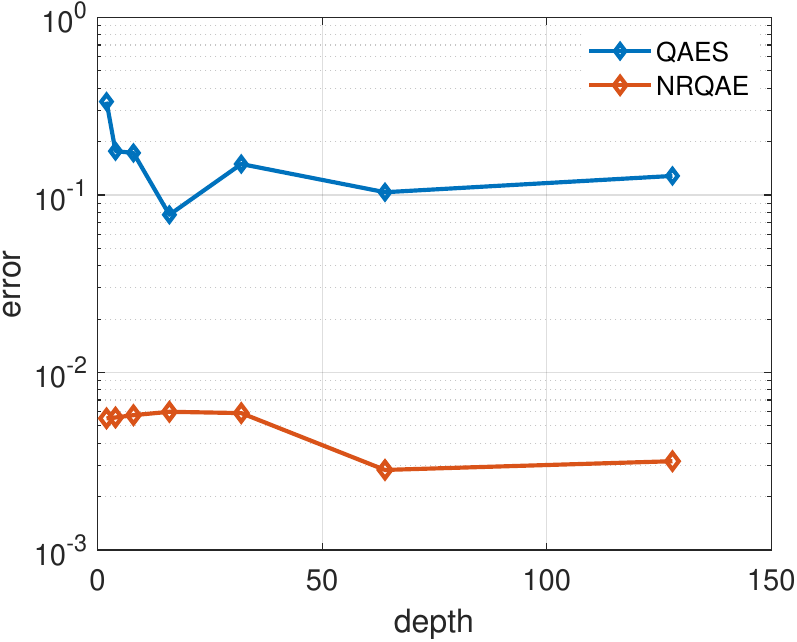}};
\node[anchor=north west] (d)
  {\includegraphics[width=4cm]{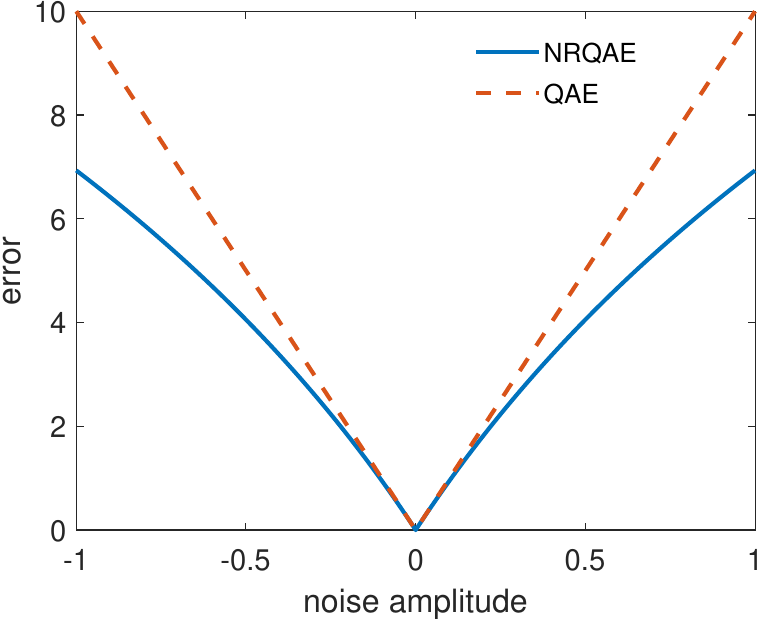}};
\endscope
\foreach \n in {a,b,c,d} {
  \node[anchor=north west] at (\n.north west) {(\n)};
}
\end{tikzpicture}
\caption{\label{fig:17}(a)The blue line shows the variation of noise with depth obtained by the existing algorithm under the influence of noise. The orange line represents the variation of NRQAE with depth. Obviously, in the case of noise, NRQAE has higher calculation accuracy. (b)The yellow line generally represents the traditional amplitude estimation, and the blue line represents the amplitude estimation of the Grover operator-based algorithm. (c)We set the depth of the quantum circuit as from 20 to 200, namely the number of the result through the above algorithm, and compare it with the accurate solution to obtain the precision of the quantum circuit. Through the final results calculated by different quantum circuits, it can be seen that with the increase of the depth of quantum circuits, the calculation accuracy has been greatly improved. (d)The yellow line is an ideal illustration of improved accuracy. The starting points of the yellow and blue lines are set the same. It can be found that the yellow and blue lines almost coincide, indicating that the calculation results are in line with expectations, that is, the calculation accuracy is improved by a multiple of $2n$. }
	\label{fig3}
\end{figure}

In our initial observations depicted in Fig.\ref{fig:17}(a), we observed rapid improvement in calculation accuracy with increasing layers. The comparison between theoretical and simulated optimization effects, as seen in Fig.\ref{fig:17}(b), confirmed their agreement. When considering noisy scenarios, our algorithm demonstrated higher accuracy even in the presence of accumulated noise with increasing depth, maintaining stability, as revealed in Fig.~\ref{fig:10}. 

\section{D: Proof about the error in $t_n$}
\begin{lemma}
Let $\lambda^{\text{i}}_1$ and $\lambda^{\text{i}}_2$ be the eigenvalues of $M_G$ in the invariant subspace $\text{V}$, corresponding to the eigenvectors $|\rho^{\text{i}}_1\rangle\rangle$ and
$|\rho^{\text{i}}_2\rangle\rangle$, respectively. Then the perturbative eigenvalues $\lambda_1$ and $\lambda_2$ are given by
\begin{align}
    &\lambda_1 - \lambda^{\text{i}}_1 = \frac{\langle\langle\rho^{\text{i}}_1 |\delta M_G|\rho^{\text{i}}_1\rangle\rangle}{\langle\langle\rho^{\text{i}}_1|\rho^{\text{i}}_1\rangle\rangle} + O(\|\delta M_G\|^2),\\
        &\lambda_2 - \lambda^{\text{i}}_2 = \frac{\langle\langle\rho^{\text{i}}_2 |\delta M_G|\rho^{\text{i}}_2\rangle\rangle}{\langle\langle\rho^{\text{i}}_2|\rho^{\text{i}}_2\rangle\rangle} + O(\|\delta M_G\|^2),
\end{align}
where $\delta M_G = NM_G - M_G$.
\end{lemma}
\textbf{Proof}: By subtracting $M_G|\rho^{\text{i}}_1\rangle\rangle =\lambda^{\text{i}}_1  |\rho^{\text{i}}_1\rangle\rangle $ **from** $NM_G|\rho_1\rangle\rangle =\lambda_1  |\rho_1\rangle\rangle$, we obtain
\begin{align}
    M_G( |\rho_1\rangle\rangle - |\rho^{\text{i}}_1\rangle\rangle) + \delta M_G |\rho_1\rangle\rangle = \lambda_1 (|\rho_1\rangle\rangle - |\rho^{\text{i}}_1\rangle\rangle) + (\lambda_1 - \lambda^{\text{i}}_1) |\rho^{\text{i}}_1\rangle\rangle.
    \label{eq: proof1}
\end{align}
Multiply $\langle\langle\rho^{\text{i}}_1|$ on the left side of Eq.~\ref{eq: proof1} to get
\begin{align}
    \langle\langle\rho^{\text{i}}_1|M_G|(|\rho_1\rangle\rangle - |\rho^{\text{i}}_1\rangle\rangle) + \langle\langle\rho^{\text{i}}_1|\delta M_G||\rho^{\text{i}}_1\rangle\rangle = \lambda^{\text{i}}_1\langle\langle\rho^{\text{i}}_1|(|\rho_1\rangle\rangle - |\rho^{\text{i}}_1\rangle\rangle) + (\lambda_1 - \lambda^{\text{i}}_1)\langle\langle\rho^{\text{i}}_1|\rho^{\text{i}}_1\rangle\rangle,
\end{align}
where we have ignored second-order terms $\delta M_G (|\rho_1\rangle\rangle - |\rho^{\text{i}}_1\rangle\rangle)$ and $(\lambda_1 - \lambda^{\text{i}}_1)(|\rho_1\rangle\rangle - |\rho^{\text{i}}_1\rangle\rangle)$. Taking into account the relationship $\langle\langle\rho^{\text{i}}_1|M_G = \lambda^{\text{i}}_1\langle\langle\rho^{\text{i}}_1| $, we can derive $\lambda_1 - \lambda^{\text{i}}_1$. Following similar process, $\lambda_2 - \lambda^{\text{i}}_2$ can be derived.

\begin{lemma}
    Suppose that the nondegenerate eigenvectors of $NM_G$ are $|\rho_1\rangle\rangle$ and $|\rho_2\rangle\rangle$. Then There exist $c_1$ and $c_2$, such that $\|c|\rho_1^{\text{i}}\rangle\rangle + c|\rho_2^{\text{i}}\rangle\rangle - c_1|\rho_1\rangle\rangle - c_2|\rho_2\rangle\rangle\|\sim O(\|\delta M_G\|)$.
\end{lemma}
\textbf{Proof}:
Multiply $\langle\langle\rho^{\text{i}}_2|$ on the left side of Eq.~\ref{eq: proof1} to get
\begin{align}
     \langle\langle\rho^{\text{i}}_2|M_G|(|\rho_1\rangle\rangle - |\rho^{\text{i}}_1\rangle\rangle) + \langle\langle\rho^{\text{i}}_2|\delta M_G||\rho_1\rangle\rangle = \lambda_1\langle\langle\rho^{\text{i}}_2|(|\rho_1\rangle\rangle - |\rho^{\text{i}}_1\rangle\rangle) + (\lambda_1 - \lambda^{\text{i}}_1)\langle\langle\rho^{\text{i}}_2|\rho^{\text{i}}_1\rangle\rangle.
\end{align}
Ignore the second-order terms, we can get
\begin{align}
    \langle\langle\rho^{\text{i}}_2|\delta M_G||\rho^{\text{i}}_1\rangle\rangle = (\lambda^{\text{i}}_1 - \lambda^{\text{i}}_2)\langle\langle\rho^{\text{i}}_2|(|\rho_1\rangle\rangle - |\rho^{\text{i}}_1\rangle\rangle).
\end{align}
Thus the difference between $|\rho_1\rangle\rangle$ and $|\rho_1^{\text{i}}\rangle\rangle$ can be written as  
\begin{align}
    |\rho_1\rangle\rangle - |\rho^{\text{i}}_1\rangle\rangle = \frac{ |\rho^{\text{i}}_2\rangle\rangle \langle\langle\rho^{\text{i}}_2|\delta M_G|\rho^{\text{i}}_1\rangle\rangle}{(\lambda^{\text{i}}_1 - \lambda^{\text{i}}_2) \langle\langle\rho^{\text{i}}_2 |\rho^{\text{i}}_2\rangle\rangle},
    \label{eq: delta_rho}
\end{align}
where we have assumed that $\langle\langle\rho^{\text{i}}_1| (|\rho_1\rangle\rangle - |\rho^{\text{i}}_1\rangle\rangle) = 0$. Furthermore, we can state that
\begin{align}
    \langle\langle\rho^{\text{i}}_2|(|\rho_1\rangle\rangle - |\rho^{\text{i}}_1\rangle\rangle) \leq \frac{\|\delta M_G\|}{\lambda^{\text{i}}_1 - \lambda^{\text{i}}_2} + O(\|\delta M_G\|^2).
\end{align}
The initial state and measurement is
\begin{align}
    |\Tilde{}{\rho}\rangle\rangle &= c|\rho^{\text{i}}_1\rangle\rangle + c|\rho^{\text{i}}_2\rangle\rangle\nonumber\\
    & = c_1 |\rho_1\rangle\rangle + c_2|\rho_2\rangle\rangle + |\Delta \rho\rangle\rangle.
    \label{eq: rho}
\end{align}
Multiply $\langle\langle\rho_1|$ on the left side of Eq.~\ref{eq: rho} to get
\begin{align}
c\langle\langle\rho_1|\rho^{\text{i}}_1\rangle\rangle + c\langle\langle\rho_1|\rho^{\text{i}}_2\rangle\rangle &= c_1\langle\langle\rho_1|\rho_1\rangle\rangle.
    \label{eq: error_c_1}
\end{align}
Similarly, 
\begin{align}
        c\langle\langle\rho_2|\rho^{\text{i}}_1\rangle\rangle + c\langle\langle\rho_2|\rho^{\text{i}}_2\rangle\rangle &= c_2\langle\langle\rho_2|\rho_2\rangle\rangle.
    \label{eq: error_c_2}
\end{align}
Thus we know that the coefficient $c_1$ is actually
\begin{align}
    c_1 & = \frac{ c\langle\langle\rho_1|\rho^{\text{i}}_1\rangle\rangle + c\langle\langle\rho_1|\rho^{\text{i}}_2\rangle\rangle}{\langle\langle\rho_1|\rho_1\rangle\rangle}\nonumber\\
    & \leq \frac{c\langle\langle\rho^{\text{i}}_1|\rho^{\text{i}}_1\rangle\rangle + c\frac{\|\delta M_G\|}{\lambda^{\text{i}}_1 - \lambda^{\text{i}}_2} + O(\|\delta M_G\|^2)}
{\langle\langle\rho^{\text{i}}_1|\rho^{\text{i}}_1\rangle\rangle +  \frac{\|\langle\langle\rho^{\text{i}}_2|\delta M_G|\rho^{\text{i}}_1\rangle\rangle\|}{\|\lambda^{\text{i}}_1 - \lambda^{\text{i}}_2 \|^2} +O(\|\delta M_G\|^2) }.
\end{align}
As a result, the coefficient $c_1$ satisfies
\begin{align}
    |c_1-c| &\leq \frac{c\frac{\|\delta M_G\|}{\lambda^{\text{i}}_1 - \lambda^{\text{i}}_2} + O(\|\delta M_G\|^2)} {\langle\langle\rho^{\text{i}}_1|\rho^{\text{i}}_1\rangle\rangle +  \frac{\|\langle\langle\rho^{\text{i}}_2|\delta M_G|\rho^{\text{i}}_1\rangle\rangle\|}{\|\lambda^{\text{i}}_1 - \lambda^{\text{i}}_2 \|^2} +O(\|\delta M_G\|^2) }\nonumber\\
    &\sim O(|\delta M_G|).
\end{align}

From Eq.~\ref{eq: delta_rho}, we know that
\begin{align}
    (\langle\langle\rho_1| - \langle\langle\rho^{\text{i}}_1|)(|\rho_1\rangle\rangle - |\rho^{\text{i}}_1\rangle\rangle ) &=\langle\langle\rho_1| \rho_1\rangle\rangle - \langle\langle\rho^{\text{i}}_1|\rho_1\rangle\rangle - \langle\langle\rho_1|\rho^{\text{i}}_1\rangle\rangle + \langle\langle\rho^{\text{i}}_1|\rho^{\text{i}}_1\rangle\rangle\nonumber\\ 
    & = \langle\langle\rho_1| \rho_1\rangle\rangle -\langle\langle\rho^{\text{i}}_1|\rho^{\text{i}}_1\rangle\rangle\nonumber\\
    & = \frac{\|\langle\langle\rho^{\text{i}}_2|\delta M_G|\rho^{\text{i}}_1\rangle\rangle\|}{\|\lambda^{\text{i}}_1 - \lambda^{\text{i}}_2 \|^2} +O(\|\delta M_G\|^2) .
    \label{eq: rho_1_and_rho_ideal}
\end{align}

\begin{theorem}
    For arbitrary $n$, the error in the estimate of $t_n$ will not exceed $O(\|\delta M_G\|)$.
\end{theorem}
\textbf{Proof}:
From Lemma 2, the square of the magnitude of the vector $|\rho\rangle\rangle$ is 
\begin{align}
    \langle\langle\Delta \rho|\Delta\rho\rangle\rangle &=  \langle\langle\Delta \rho|\rho\rangle\rangle\nonumber\\
    & = |c|^2 ( \langle\langle\rho^{\text{i}}_1|\rho^{\text{i}}_1\rangle\rangle + \langle\langle\rho^{\text{i}}_2|\rho^{\text{i}}_2\rangle\rangle ) - c^{*} (c_1 \langle\langle\rho^{\text{i}}_1|\rho_1\rangle\rangle + c_2\langle\langle\rho^{\text{i}}_1|\rho_2\rangle\rangle )   -c^{*} (c_1 \langle\langle\rho^{\text{i}}_2|\rho_1\rangle\rangle + c_2\langle\langle\rho^{\text{i}}_2|\rho_2\rangle\rangle )\nonumber\\
    & = (|c|^2 - c^{*}c_1)\langle\langle\rho^{\text{i}}_1|\rho^{\text{i}}_1\rangle\rangle + (|c|^2 - c^{*}c_2)\langle\langle\rho^{\text{i}}_2|\rho^{\text{i}}_2\rangle\rangle - c^{*}c_2\langle\langle\rho^{\text{i}}_1|( \rho_2\rangle\rangle -\rho^{\text{i}}_2\rangle\rangle  )  - c^{*}c_1\langle\langle\rho^{\text{i}}_2|( \rho_1\rangle\rangle -\rho^{\text{i}}_1\rangle\rangle  )\nonumber\\
    &\sim O(\|\delta M_G\|).
\end{align}
As a result, the difference between $t_n$ and $|c_1| p^n \cos(2n\theta)$ will be
\begin{align}
    &\langle\langle\Tilde{\rho}|(NM_G)^n|\Tilde{\rho}\rangle\rangle - (c_1\langle\langle\rho_1| + c_2\langle\langle\rho_1|)(NM_G)^n(c_1|\rho_1\rangle\rangle + c_2|\rho_2\rangle\rangle)\nonumber\\
    =&  \langle\langle\Delta \rho|(NM_G)^n|\rho\rangle\rangle + \langle\langle\rho|(NM_G)^n|\Delta\rho\rangle\rangle + \langle\langle\Delta \rho|(NM_G)^n|\Delta\rho\rangle\rangle \nonumber\\
    \leq & \langle\langle\Delta \rho|\rho\rangle\rangle + \langle\langle\rho|\Delta\rho\rangle\rangle + \langle\langle\Delta \rho|\Delta\rho\rangle\rangle\nonumber\\
    \sim & O(\|\delta M_G\|).
\end{align}
\end{appendices}

\end{document}